# Ultrafast Studies of Hot-Hole Dynamics in Au/p-GaN Heterostructures

Giulia Tagliabue[1,2,†], Joseph S. DuChene[1,2,†], Mohamed Abdellah[3,4,†], Adela Habib[5], Yocefu Hattori[3], Kaibo Zheng[6,7], Sophie E. Canton[8,9], David J. Gosztola[10], Wen-Hui Cheng[1,2], Ravishankar Sundararaman[5], Jacinto Sá[3,11]*, Harry A. Atwater[1,2]*

[1]Thomas J. Watson Laboratory of Applied Physics, California Institute of Technology, Pasadena, California 91125 United States.
[2]Joint Center for Artificial Photosynthesis, California Institute of Technology, Pasadena, California 91125 United States.
[3]Department of Chemistry-Ångström Laboratory, Uppsala University, 75120 Uppsala, Sweden.
[4]Department of Chemistry, Qena Faculty of Science, South Valley University, 83523 Qena, Egypt.
[5]Department of Materials Science and Engineering, Rensselaer Polytechnic Institute, 110 8th Street, Troy, New York 12180, United States.
[6]Department of Chemistry, Technical University of Denmark, DK-2800 Kongens Lyngby, Denmark.
[7]Department of Chemical Physics and NanoLund, Lund University, Box 124, 22100, Lund, Sweden.
[8]ELI-ALPS, ELI-HU Non-Profit Ltd., Dugonicster 13, Szeged 6720, Hungary.
[9]Attoscience Group, Deutsche Elektronen Synchrotron (DESY), Notkestrasse 85, D-22607 Hamburg, Germany.
[10]Center for Nanoscale Materials, Nanoscience and Technology Division, Argonne National Laboratory, Argonne, Illinois 60439, United States.
[11]Institute of Physical Chemistry, Polish Academy of Sciences, Warsaw 01-224 Poland.

[†]These authors contributed equally

**Abstract**

Harvesting non-equilibrium "hot" carriers from photo-excited metal nanoparticles has enabled plasmon-driven photochemical transformations and tunable photodetection with resonant nanoantennas[1-13]. Despite numerous studies on the ultrafast dynamics of hot electrons[14-26], to date, the temporal evolution of hot holes in metal-semiconductor heterostructures remains unknown. An improved understanding of the carrier dynamics in hot-hole-driven systems is needed to help expand the scope of hot-carrier optoelectronics beyond hot-electron-based devices. Here, using ultrafast transient absorption spectroscopy, we show that plasmon-induced hot-hole injection from gold (Au) nanoparticles into the valence band of p-type gallium nitride (p-GaN) occurs within 200 fs, placing hot-hole transfer on a similar timescale as hot-electron transfer[22,25]. We further observed that the removal of hot holes from below the Au Fermi level exerts a discernible influence on the thermalization of hot electrons above it, reducing the peak electronic temperature and decreasing the electron-phonon coupling time relative to Au samples without a pathway for hot-hole collection. First principles calculations[27-29] corroborate these experimental observations, suggesting that hot-hole injection modifies the relaxation dynamics of hot electrons in Au nanoparticles through ultrafast modulation of the *d*-band electronic structure. Taken together, these ultrafast studies substantially advance our understanding of the temporal evolution of hot holes in metal-semiconductor heterostructures and suggest new strategies for manipulating and controlling the energy distributions of hot carriers on ultrafast timescales.

The absorption of light by metallic nanostructures, either via interband or intraband optical transitions, generates a non-equilibrium distribution of "hot" electrons and holes (Figure 1a) whose energetics above and below the metal Fermi level are collectively controlled by both the incident photon energy and the metal band structure[5,27,28]. These non-thermal "hot" carriers quickly equilibrate via electron-electron scattering processes on an ultrafast timescale ($t$ ~fs–ps) to establish an elevated electronic temperature ($T_e$) relative to the lattice temperature ($T_l$). The excited hot-carrier distribution subsequently equilibrates with the underlying lattice via electron-phonon coupling ($t$ ~ps–ns) as the nanostructure dissipates heat to its local surroundings ($t$ > ns). Despite numerous studies of hot-carrier dynamics, the vast majority of reports involve isolated colloidal metal nanoparticles suspended in aqueous solution[15-19,21-24]. However, the realization of hot-carrier optoelectronics, however, requires placing metal nanostructures onto semiconductor substrates to facilitate carrier injection and transport in optoelectronic devices. Indeed, the collection of hot carriers is typically accomplished through the formation of a rectifying interfacial Schottky barrier ($\Phi_B$) contact with n-type (p-type) semiconductors to quickly capture hot electrons (hot holes) prior to their thermalization with the phonon bath ($t$ ~ 1 ps) (Figure 1b,c)[1-3,5-11]. While the vast majority of metal-semiconductor heterostructures have been devised to enable hot-electron collection[1-3,5-11], few systems suitable for hot-hole capture and conversion have been reported[30,31]. Recent *ab initio* calculations[27-29] suggest that the removal of hot holes from the metal *d*-bands via injection to a p-type semiconductor could significantly alter the relaxation dynamics of photo-excited hot electrons within the *sp*-band of the metal[29], offering new opportunities for manipulating the energy distributions of hot carriers on metal nanostructures. Accordingly, the ultrafast dynamics of hot holes at metal/p-type semiconductor interfaces must be understood to evaluate the potential implications for device applications.

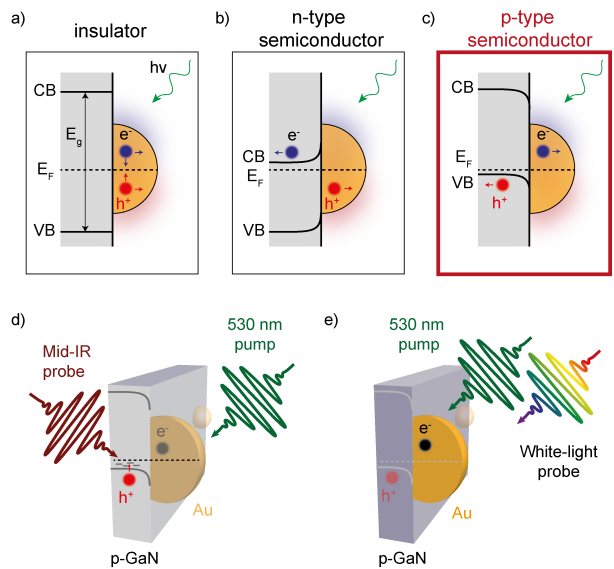

*Figure 1. Plasmon-induced hot carriers in metal nanostructures. Plasmon-induced hot electron-hole pair production in Au nanoparticles supported on a) insulating substrate, b) n-type semiconductor, or c) p-type semiconductor, depicting the role of the substrate in preventing (a) or facilitating (b,c) interfacial charge separation of hot electron-hole (e-h) pairs at the metal-semiconductor heterojunction. The energies of the conduction band (CB), valence band (VB), band gap ($E_g$), and Fermi level ($E_F$) are shown. In all cases, the incident photon energy ($hv$) that excites hot carriers on the metal is less than $E_g$ of the support. d,e) Experimental pump-probe design using 530 nm pump wavelength to create hot electron-hole pairs on the Au nanoparticles and initiate hot-hole injection to the p-type GaN support while (d) probing the hot holes within the p-GaN valence band via infrared transient absorption spectroscopy, or (e) probing the hot electron dynamics on the Au nanoparticles with a broadband white-light source across the visible regime.*

A useful platform for the study of photo-excited hot holes consists of gold nanoparticles dispersed onto p-type gallium nitride (p-GaN). A sizable Schottky barrier ($\Phi_B$ = 1.1 eV) is established across the Au/p-GaN interface which ensures that only hot holes with energies in excess of 1 eV below the Au Fermi level can be collected by the p-GaN support[30]. The wide band gap of p-GaN ($E_G$ = 3.4 eV) excludes visible-light excitation of free carriers directly within the semiconductor support and thereby serves exclusively to capture optically-excited hot holes from the Au nanoparticles[30]. Such a system enables spectrally distinguishing the dynamics of hot holes within the valence band of p-GaN, which can be probed via free-carrier absorption in the infrared regime (Figure 1d), from hot electrons in the metal, which are separately monitored across the visible spectrum with a broadband white-light probe (Figure 1e). Plasmonic Au nanoparticles 7.3 ± 2.4 nm in diameter were uniformly distributed across a p-type GaN epi-film on sapphire via electron-beam deposition (Figure 2a and Figure S1). No intervening adhesion layer was used to form the Au/p-GaN interface (see Methods). The absorption spectrum of Au/p-GaN heterostructures exhibits a peak in the visible region at 568 nm, attributable to the surface plasmon resonance of Au nanoparticles (Figure 2a). Fringes present in the spectra are due to Fabry-Pérot interference within the p-GaN film itself (Figure S2).

Plasmon-induced hot-hole dynamics in Au/p-GaN heterostructures were monitored via ultrafast transient absorption spectroscopy (see Methods). Optical excitation of hot carriers in the Au nanoparticle array was initiated with a 530 nm pump pulse ($\lambda_{pump}$) and the temporal evolution of hot holes within the p-GaN valence band were probed across the infrared regime (Figure S3). As indicated by the steep rise in transient absorption (ΔAbs) at a probe wavelength of 4.85 µm, plasmon excitation induces hot-hole injection from Au nanoparticles into the p-GaN valence band within the 200 fs temporal resolution of our experimental setup (Figure 2b). Significantly, this observation confirms that hot-hole injection in Au/p-GaN heterostructures occurs on a similar timescale as previously reported for hot-electron injection at the Au/TiO$_2$ interface.[22,25] The relaxation dynamics of hot holes can be fit to a multiexponential function, which exhibits a fast decay component ($\tau_1$) on the ~1 ps timescale commensurate with electron-phonon coupling, followed by two slower components; one occurs on the tens of picoseconds timescale ($\tau_2$ ~ 10 ps) and another on the order of a few nanoseconds ($\tau_3$ ~ 5 ns). As a control, no transient absorption signal was observed from bare p-GaN supports even under increased pump power (Figure 2b, open circles). A linear relationship between the incident pump power and the transient absorption signal obtained from Au/p-GaN indicates that these observations are not the result of non-linear optical processes (Figure 2b, black points and Figure S4). We also note that excitation below the interband threshold of Au ($\lambda_{pump}$ = 750 nm) fails to elicit any transient absorption signal, confirming that interband excitation is responsible for hot-hole injection to the p-GaN valence band (Figure S5).

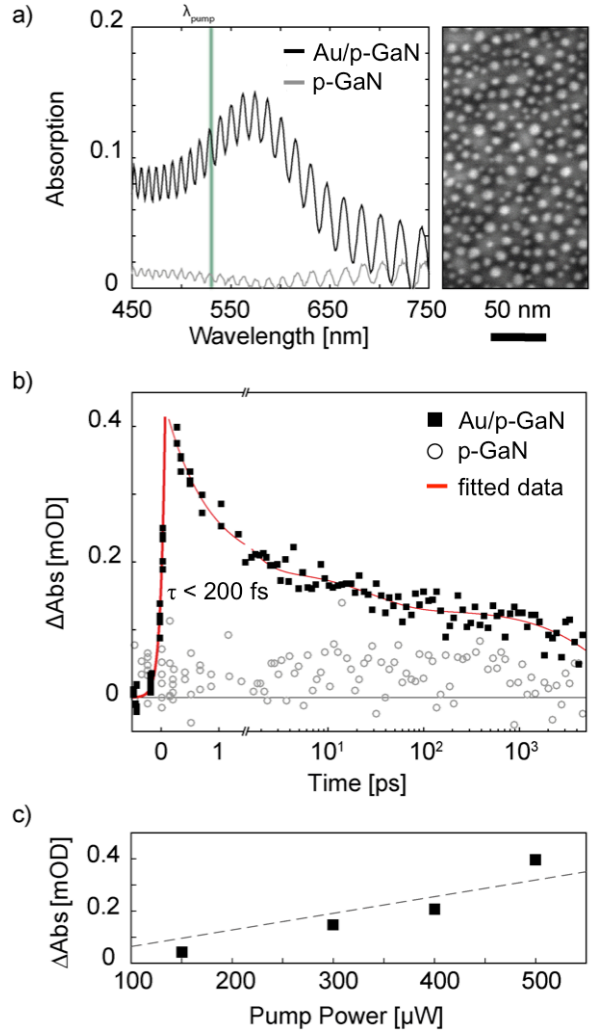

*Figure 2. Infrared transient absorption spectroscopy of hot-hole dynamics in Au/p-GaN heterostructures. a) Absorption spectra of Au/p-GaN (black curve) and bare p-GaN substrate (grey curve) along with the corresponding SEM image of Au nanoparticles (mean diameter 7.3 ± 2.4 nm) on p-GaN support. The green line in the absorption spectra indicates the pump laser wavelength $\lambda_{pump}$ = 530 nm used for plasmon excitation of Au nanoparticles. b) Ultrafast transient rise and decay probed at 2060 $cm^{-1}$ ($\lambda_{probe}$ = 4.85 µm) obtained from Au/p-GaN (black points) and bare p-GaN (open circles) upon 530 nm pump pulse at an incident power of 500 µW and 750 µW respectively. The red line shows a fit to the experimental data, which exhibits an instrument-limited rise time ($\tau$) of less than 200 fs. c) Power-dependent transient absorption peak signal monitored at 2060 $cm^{-1}$ (4.85 µm) from Au/p-GaN heterostructures displaying linear behavior across the range from 150 µW to 500 µW.*

The influence of ultrafast hot-hole collection on the dynamics of hot electrons was then evaluated by comparing the differential absorption spectrum of Au nanoparticles on p-GaN with insulating glass ($SiO_2$) supports. As charge separation is prohibited across the Au/$SiO_2$ interface[32], this inert support serves as reference for the hot-electron dynamics observed on p-GaN. The Au nanoparticles on $SiO_2$ exhibit a mean diameter of 14.7 ± 5.7 nm and display a surface plasmon resonance peak centered around 546 nm (Figure S6). Although larger on $SiO_2$ than on p-GaN, Au nanoparticles in this regime (d = 7–15 nm) do not exhibit any size-dependent carrier dynamics[14-19]. Plasmon excitation was initiated with a 530 nm pump pulse while probing the ultrafast dynamics of hot electrons on the Au nanoparticles across the visible regime ($\lambda_{probe}$ = 450–750 nm). Figure 3 shows the temporal evolution of the differential absorption (ΔAbs) spectrum as a two-dimensional color map from Au/p-GaN (Figure 3a) and Au/SiO$_2$ (Figure 3b). The spectral profile exhibits a transient bleach of the surface plasmon resonance (blue-shaded region) accompanied by two transient absorption "wings" (red-shaded regions) on either side of this central feature. The transient bleach of the surface plasmon band from Au/p-GaN and Au/SiO$_2$ is plotted in the time domain in Figure 3c,d. The relaxation dynamics are primarily governed by electron-phonon coupling ($\tau_{e-ph}$) processes occurring on a picosecond timescale, with a decay time of around 2.1 ps on Au/p-GaN (Figure 3c) and around 3 ps on Au/SiO$_2$ (Figure 3d).

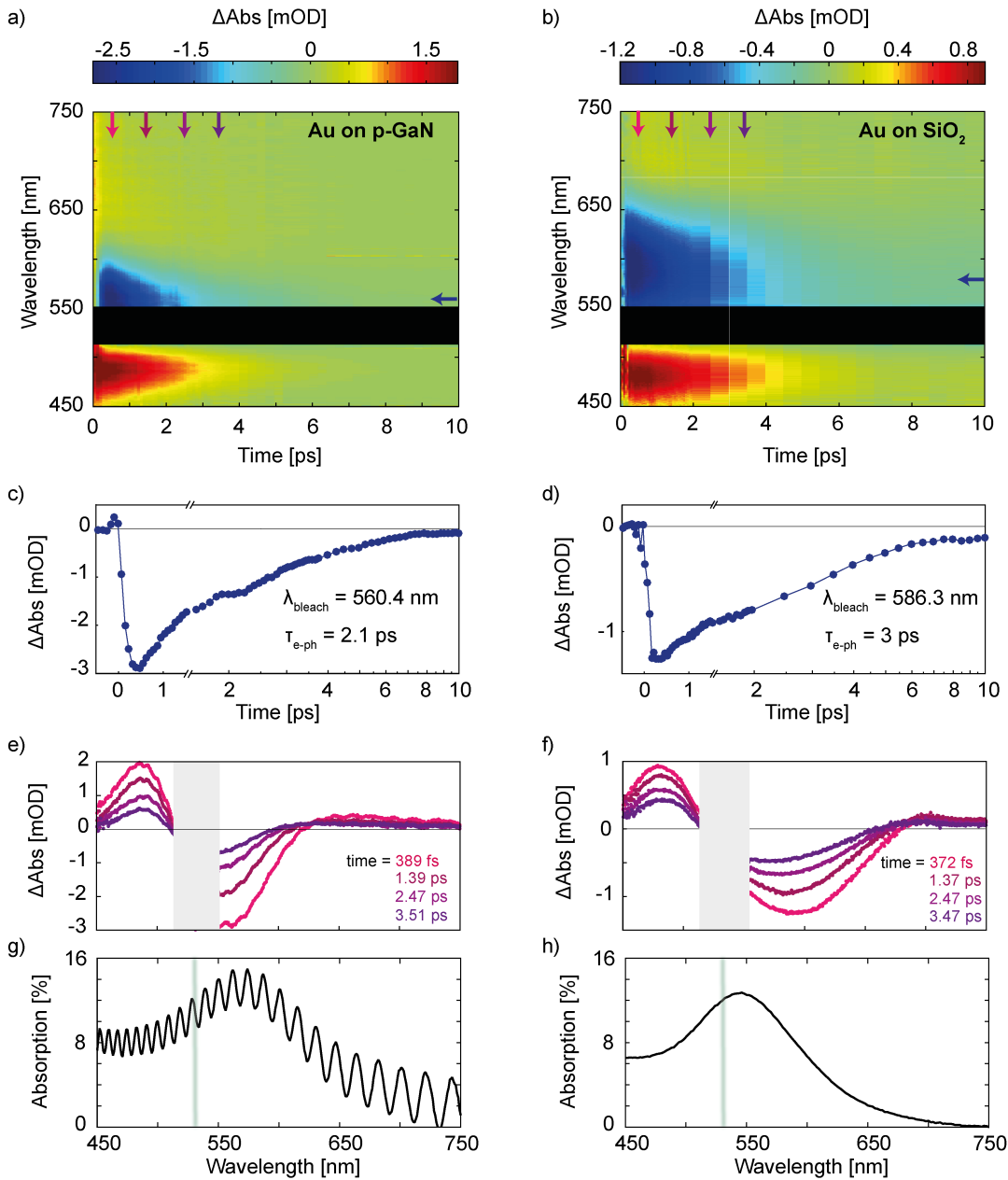

*Figure 3. Visible-light transient absorption spectroscopy of hot-electron dynamics in Au/p-GaN and Au/SiO$_2$ heterostructures. a,b)* Two-dimensional color maps of the temporal evolution of the surface plasmon resonance of Au/p-GaN (a) and Au/SiO$_2$ (b). The black box denotes the spectral region obscured by the pump pulse, while arrows indicate the location of the temporal (purple arrows)

*and spectral (blue arrows) cuts shown in c-f. **c,d)** Spectral cuts at 560 nm for Au/p-GaN (c) and 586 nm for Au/SiO$_2$ (d), showing the time dynamics of the transient bleach. **e,f)** Temporal cuts taken at various delay times (t ~ 400 fs, 1.4 ps, 2.5 ps, 3.5 ps) with respect to the pump pulse showing the recovery dynamics of the transient absorption spectrum from Au/p-GaN (e) and Au/SiO$_2$ (f). The grey box denotes the spectral region obscured by the pump pulse. **g,h)** steady-state absorption spectrum from Au/p-GaN (g) and Au/SiO$_2$ (h) shown for comparison. The green line indicates the laser excitation wavelength of 530 nm.*

To visualize changes in the spectral response of the carrier dynamics, we plot several cuts of the differential absorption spectrum at various delay times from both systems in Figure 3e,f. For Au/SiO$_2$, the peak of the transient bleach is initially red-shifted to around 580 nm relative to the surface plasmon band at 546 nm and steadily blue-shifts back towards the plasmon resonance over several picoseconds as the electron gas cools via electron-phonon coupling (Figure 3f). The features observed from Au/SiO$_2$ are generally consistent with prior ultrafast studies of hot-carrier dynamics in metal nanoparticles[14-19]. However, the carrier dynamics of the Au nanoparticles differ substantially on the p-GaN support; the peak position of the transient bleach in Au/p-GaN is initially blue-shifted to 560 nm relative to the plasmon resonance at 568 nm, and exhibits no apparent spectral shift in either direction as the differential absorption decays back towards the ground state (Figure 3e). Furthermore, the transient bleach is much more spectrally narrow on p-GaN and the overall shape of the differential absorption appears less symmetric with respect to the central feature (Figure 3e). Such pronounced differences between Au/p-GaN and Au/SiO$_2$ suggest that ultrafast hot-hole collection by the underlying p-GaN support significantly affects the hot-electron dynamics on the Au nanoparticles.

The width of the transient bleach from plasmonic-metal nanoparticles is known to directly correlate with the elevated electronic temperature ($T_e$) of the system established after photon absorption[17,18]. Indeed, the overall width of the transient bleach from both Au/p-GaN and Au/SiO$_2$ grew with increasing pump power (Figure S7 and Figure S8). The narrower width of the transient bleach between these two heterostructures indicates that hot electrons attain a lower $T_e$ in Au/p-GaN than in Au/SiO$_2$. Given that the electron-phonon coupling time ($\tau_{e-ph}$) in Au nanoparticles is proportional to the electronic temperature $T_e$[14-19], we experimentally evaluated $\tau_{e-ph}$ for the Au/p-GaN and Au/SiO$_2$ heterostructures by fitting the decay time of the transient bleach for different incident pump powers. The determined values of $\tau_{e-ph}$ are plotted as a function of the absorbed energy density $U_{Abs}$ (J m$^{-3}$) to account for the differences in nanoparticle size, optical density, and substrate coverage between Au/p-GaN and Au/SiO$_2$ heterostructures (see Methods). As shown in Figure 4a, Au nanoparticles supported on p-GaN (black squares) exhibit shorter $\tau_{e-ph}$ at all pump powers with a much smaller slope than those supported on SiO$_2$ (blue triangles). Extrapolating both curves to zero incident power yields a value of $\tau_{e-ph}$ ~1 ps for each system, consistent with prior reports of colloidal Au nanoparticles[14-19]. To confirm that this difference in $\tau_{e-ph}$ is not solely attributable to differences in thermal conductivity of the two substrates, Au nanoparticles were also deposited onto sapphire (Al$_2$O$_3$) supports (Figure S9). Despite the nearly 30 times higher thermal conductivity of Al$_2$O$_3$ than SiO$_2$, the slope of $\tau_{e-ph}$ is nearly identical for Au/SiO$_2$ and Au/Al$_2$O$_3$ (Figure S10), confirming that the differences between Au/SiO$_2$ and Au/p-GaN cannot be explained simply by differences in thermal conductivity. Thus, these results imply that ultrafast hot-hole collection by p-GaN alters the thermalization dynamics of hot electrons on the Au nanoparticles and lowers the electronic temperature $T_e$ attained by Au/p-GaN.

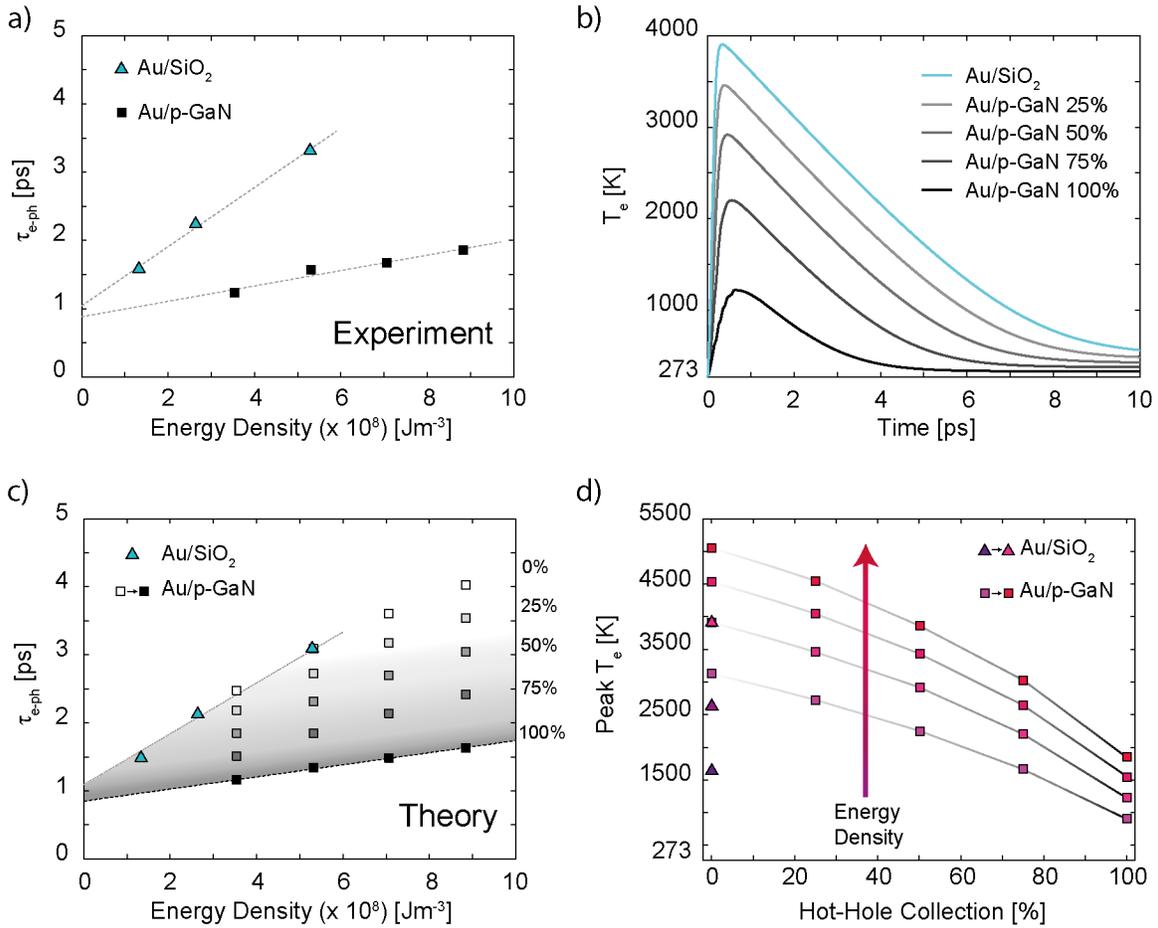

*Figure 4. Influence of ultrafast hot-hole collection on the dynamics of hot electrons in Au nanoparticles. a) Experimental plot of electron-phonon coupling time ($\tau_{e-ph}$) as a function of absorbed laser power for Au/SiO$_2$ (blue triangles) and Au/p-GaN (black squares) – dashed lines serve as guides to the eye. b) Theoretical calculation showing the temporal evolution of the electronic temperature ($T_e$) after laser excitation at 530 nm ($U_{abs}$ = 5.3x10$^8$ Jm$^{-3}$) in Au/SiO$_2$ (blue curve) and Au/p-GaN (grey curves). For Au/p-GaN, the influence of removing hot holes 1 eV below the Au Fermi level is shown (light-grey curve: 25% → black curve: 100% hot-hole collection) c) Theoretical calculation of electron-phonon coupling time ($\tau_{e-ph}$) as a function of absorbed laser power for Au/SiO$_2$ (blue triangles) and Au/p-GaN (grey squares). For Au/p-GaN, the influence of removing hot holes 1 eV below the Au Fermi level is shown (white squares : 0% → black squares: 100% hot-hole collection). d) Theoretical calculation of the peak electronic temperature (Te) established in Au/SiO$_2$ (triangles) and Au/p-GaN (squares) for various absorbed energy densities plotted as a function of the total percentage of hot holes 1 eV below the Au Fermi level that are collected from the Au nanoparticles.*

The electronic heat capacity $C_e$ of Au nanoparticles, which is proportional to both the electronic temperature $T_e$ and the electron density $N_e$, is highly sensitive to the electronic structure of the Au $d$-bands[29]. In particular, a dramatic increase in $C_e(T_e,N_e)$ occurs around 1.8 eV below the Au Fermi level due to the large density of $d$-band states that become accessible at these energies[29]. Ultrafast hot-hole injection to the p-GaN valence band effectively raises the electron density $N_e$ of the Au nanoparticles on timescales commensurate with electron-electron thermalization, thereby increasing $C_e(T_e,N_e)$ of the metal as the hot electrons are actively establishing an elevated electronic temperature $T_e$ above the Au Fermi level. Since the electronic density of states and corresponding electronic heat capacity $C_e(T_e,N_e)$ determine the peak electronic temperature $T_e$ attained via electron-electron scattering, we employ *ab initio* electronic structure calculations[27-29] to further examine the influence of hot-hole injection on the thermalization dynamics of hot electrons in the Au nanoparticles (see Methods). As shown in Figure 4b, removing hot holes from the Au *d*-bands modifies the thermalization dynamics of hot electrons in the Au nanoparticles and limits the maximum $T_e$ achieved by the electron gas in Au/p-GaN (grey curves) relative to Au/SiO$_2$ (blue curve). This lower $T_e$ in Au/p-GaN is consistent with the observed difference in $\tau_{e\text{-ph}}$ between these two heterostructures (Figure 4a). Theoretical estimation of $\tau_{e\text{-ph}}$ as a function of the percentage of hot-hole removal at several different absorbed energy densities $U_{Abs}$ further supports the interpretation of the experimentally observed trend: increased hot-hole collection leads to a decrease in $\tau_{e\text{-ph}}$, and therefore a lower $T_e$ of the electron gas (Figure 4c). The excellent quantitative agreement between the experimentally determined value of $\tau_{e\text{-ph}}$ (Figure 4c) and that predicted from theory (Figure 4c) would suggest that the vast majority (~75%) of the hot holes that reach the Au/p-GaN interface with energies in excess of the Schottky barrier height ($\Phi_B$ = 1 eV) are injected into the p-GaN valence band. Figure 4d shows the peak $T_e$ established after photon absorption plotted as a function of the total percentage of hot holes that are removed from the Au nanoparticles for various absorbed energy densities. In Au, plasmon decay deposits most of the energy in hot holes that are more than 2 eV below the Fermi level[28]. Consequently, removing 75% of hot holes 1 eV below the Fermi level eliminates a significant fraction of the absorbed energy, reducing the peak $T_e$ by ~50% relative to the $T_e$ that would be expected without any hot-hole removal (Figure 4d). Note, however, that at the typical absorbed energy densities, the net change in the conduction electron density $\Delta N_e$ in Au is only ~1% because the number of photo-excited carriers is still a small fraction of the overall electron density $N_e$ (Figure S11). Taken together, the close correspondence between experiment and theory confirms that the observed differences in both spectral and temporal dynamics between Au/p-GaN and Au/SiO$_2$ are attributable to the ultrafast ($t$ < 200 fs) injection of hot holes from Au nanoparticles to the valence band of p-GaN.

In summary, we demonstrate that the relaxation dynamics of hot electrons in Au nanoparticles are considerably altered by selectively extracting hot holes from below the Au Fermi level on an ultrafast timescale. Our observation of ultrafast hot-hole injection ($t$ ~fs) coupled with the relatively long lifetime of the charge-separated state ($t$ ~ns) further suggests that hot-hole-based optoelectronics could offer comparable device performance to that obtained with hot-electron-based systems. The results of these ultrafast studies suggest new strategies for manipulating the ultrafast dynamics of hot electrons in photocatalytic systems and optoelectronic devices by controlling the collection of hot holes with p-type semiconductors.

**Author Contributions**
J.S.D., G.T., and H.A.A. conceived the idea, designed the experiments, and wrote the manuscript with contributions from all authors. M.Q., Y.H., and J.S. performed infrared transient absorption spectroscopy experiments. M.Q., K.Z., S.E.C., and D.J.G. performed visible transient absorption spectroscopy experiments. A.H. and R.S. performed theoretical calculations. J.S.D. and G.T. fabricated and characterized materials. W.-H.C. acquired absorption spectra of materials. H.A.A. supervised the project. All authors have given approval to the final version of the manuscript. ‡These authors contributed equally.

**Acknowledgements**
This material is based upon work performed by the Joint Center for Artificial Photosynthesis, a DOE Energy Innovation Hub, supported through the Office of Science of the U.S. Department of Energy under Award No. DE-SC0004993. A portion of the ultrafast spectroscopy work was performed at the Center for Nanoscale Materials, a U.S. Department of Energy Office of Science User Facility, and supported by the U.S. Department of Energy, Office of Science, under Contract No. DE-AC02-06CH11357. G.T. acknowledges support from the Swiss National Science Foundation through the Early Postdoc Mobility Fellowship, grant n. P2EZP2_159101 and the Advanced Mobility Fellowship, grant n. P300P2_171417.

**Methods**
**Fabrication of Au/p-GaN and Au/SiO$_2$ heterostructures.** Plasmonic Au/p-GaN photocathodes were constructed via evaporation of Au thin-films onto commercial p-type GaN on sapphire substrates (Pam-Xiamen). Immediately prior to Au evaporation, the p-GaN substrates were pre-treated with dilute $NH_4OH$ solution (0.02% v/v%) for 30 s to remove any native oxide, followed by 30 s of copious washing in Nanopure™ water. The p-GaN/sapphire substrate was then blown dry with $N_2$ gas and loaded into the vacuum chamber. A 1.5 nm-thick film of Au was deposited onto the p-GaN surface using electron-beam physical vapor deposition at a base pressure of ca. $1 \times 10^{-7}$ torr and a deposition rate of 1.0 Å s$^{-1}$. The Au/p-GaN films were then annealed in ambient air at 300 °C for 1 h to ensure coalescence of the discontinuous Au thin-film into Au nanoparticles and achieve good adhesion with the underlying p-GaN surface. Note that there is no interfacial adhesion layer used to construct the Au/p-GaN heterojunction. The Au/SiO$_2$ films were prepared the same way as the Au/p-GaN films described above, except that glass was used as the substrate. The glass substrates (VWR soda-lime glass) were sequentially cleaned with acetone, isopropanol, and then water for 5 min each while sonicating in a water bath. The substrates were then copiously rinsed with water and blown dry with N$_2$ gas before loading into the vacuum chamber for Au deposition. As with Au/p-GaN, no interfacial adhesion layer was used for Au/SiO$_2$ heterostructures.

**Ultrafast transient absorption spectroscopy experiments.** The detection of plasmon-induced hot-hole injection to the p-GaN valence band was monitored via transient infrared absorption spectroscopy (TIRAS). TIRAS experiments were carried out in a femtosecond transient absorption spectrometer (Helios IR, Ultrafast Systems LLC) at room temperature. Briefly, the output of a Ti:Sapphire amplifier with integrated oscillator and pump lasers (800 nm, 40 fs, 3 kHz, Libra LHE, Coherent Inc.) was split into two beams which were used to pump two TOPAS Prime optical parametric amplifiers (OPAs) coupled with frequency mixers (Light Conversion Ltd). This setup produced a depolarized visible pump pulse ($\lambda_{pump}$ = 530 nm) and a broad mid-IR probe spectrum ($\lambda_{probe}$ = 1850–2200 cm$^{-1}$). Pump pulse energies were adjusted using a neutral density filter placed prior to the sample. The laser spot size for the 530 nm pump was ~120 μm X 160 μm. Prior to reaching the sample, the probe beam was split into equal intensity probe and reference beams. The detection of probe and reference beams was done using a femtosecond transient absorption spectrometer (Helios IR, Ultrafast Systems LLC). The instrument response function for the experiments was approximately 200 fs. All the films were measured in Specac cell. The samples were moved manually during the measurements to minimize possible laser-induced sample damage.

The hot-electron dynamics on the Au nanoparticles were monitored with femtosecond transient absorption spectroscopy. Measurements were performed at the Center for Nanoscale Materials (CNM), Argonne National Lab, USA using an amplified Ti:Sapphire laser system (Spectra Physics, Spitfire Pro) and an automated data acquisition system (Ultrafast Systems, HELIOS). The output of the amplified laser (800 nm, 150 fs, 5 kHz) was split 90/10 with the majority used to pump an optical parametric amplifier (Light Conversion, TOPAS), which provided the pump beam ($\lambda_{pump}$ = 530 nm). The remaining 10% was used to make the probe beam ($\lambda_{probe}$ = 450 – 750 nm) after traversing an optical delay line by focusing into a 2 mm thick sapphire window. The probe continuum was passed through a 785 nm short-pass filter (Semrock) to remove residual 800 nm and longer wavelengths of light. The pump and probe beams were focused (~180 μm dia. FWHM) and overlapped on the sample. Pump energies ranged from 100 μW to 600 μW (0.16–0.96 mJ cm$^{-2}$). Four scans covering 50 ps were collected and averaged with no sample degradation detected.

*Ab initio* **theoretical calculations.** We predict the evolution of electron temperature using first-principles calculations of carrier excitation by the pump pulse, electron thermalization by electron-electron (e-e) scattering as well as relaxation by electron-phonon (e-ph) coupling. We perform density-functional theory calculations of electron and phonon states, and their interaction matrix elements in the open-source JDFTx software[33]. Using Fermi's golden rule in an efficient Wannier representation[27,28], we then calculate the initial hot carrier distributions as well as the collision integrals necessary for solving the Boltzmann transport equation for the evolution of the electron distribution[21]. Briefly, the initial carrier distribution following excitation by the pump is

$$f(\epsilon, t = 0) = f_0(\epsilon) + U \frac{P(\epsilon, \hbar\omega, P_{inj})}{g(\epsilon)},$$

where $f_0$ is the Fermi distribution at equilibrium, $U$ is the pump energy absorbed density, $g(\epsilon)$ is the electronic density of states, and $P(\epsilon, \hbar\omega, p_{inj})$ is the energy distribution of carriers excited by a photon energy of $\hbar\omega$. Here, $p_{inj}$ is the percentage of holes greater than 1 eV below the Fermi level that are injected into the substrate (only for the GaN case). The absorbed pump energy density $U$ is determined by the nanoparticle size, beam diameter, pump-pulse energy and the steady-state absorption coefficient.

Evolution of the carrier distribution with time *t* is then captured by the nonlinear Boltzman equation:

$$\frac{d}{dt}f(\epsilon,t) = \Gamma_{\text{e-e}}[f](\epsilon) + \Gamma_{\text{e-ph}}[f,T_l](\epsilon),$$

where $\Gamma_{\text{e-e}}$ and $\Gamma_{\text{e-ph}}$ are the collision integrals for e-e and e-ph scattering, determined fully from first principles exactly as detailed in[21]. From the predicted distribution function $f$ at each time $t$, we estimate the electron temperature from the derivative of $f$ at the Fermi energy, because this correlates well with the dominant change in the optical response due to a large number of lower energy carriers near the Fermi level[29]. Finally, we predict the electron-phonon relaxation time, $\tau_{\text{e-ph}}$, from the fall time of the calculated electron temperature, which is in excellent agreement with fall times extracted from experiment.

**Calculation of absorbed power density.** The Au/p-GaN and Au/SiO$_2$ heterostructures exhibit different particle diameter, substrate area coverage, and optical density (i.e. steady-state absorption). Although the electron-phonon coupling constant $\tau_{\text{e-ph}}$ is known to be independent of size for metal nanoparticles in the size regime studied herein ($d$ = 7–15 nm)[14-18], a fair comparison between the two cases can only be obtained on the basis of the energy absorbed per unit volume of gold, $U_{\text{abs}}$ [Jm$^{-3}$]. We determine $U_{\text{abs}}$ according to the following expression:

$$U_{abs}[Jm^{-3}] = \frac{\text{energy absorbed per nanoparticle}}{\text{nanoparticle volume}} = \frac{\left(\frac{Abs_{530} \cdot E_{pulse}}{\pi r_{beam}^2} \cdot \frac{\pi r_{NP}^2}{C}\right)}{\frac{4}{3}\pi r_{NP}^3}$$

where $Abs_{530}$ is the steady-state absorption at 530 nm (11.7% for Au/p-GaN and 13.1% for Au/SiO$_2$), $E_{\text{pulse}}$ is the pump pulse energy (0.4– 2.0x10$^{-7}$ J/pulse for 100–500 µW), $2r_{\text{NP}}$ is the nanoparticle diameter (7.3 nm for Au/p-GaN and 14.7 nm for Au/SiO$_2$), $C$ is the nanoparticle area coverage (18.7% for Au/p-GaN and 13.7% for Au/SiO$_2$), and $2r_{\text{beam}}$ is the pump beam diameter (~180 µm FWHM).